\documentclass[journal]{IEEEtran}

\usepackage[utf8]{inputenc}
\usepackage{amsmath, amsfonts, amsthm, amssymb}
\usepackage[ruled,linesnumbered]{algorithm2e}
\usepackage{graphicx}
\usepackage{cite}
\usepackage{url}
\usepackage{textcomp}
\usepackage{xcolor}
\usepackage{algorithmic}

\usepackage{stfloats} 
\usepackage{array}
\usepackage[caption=false,font=normalsize,labelfont=sf,textfont=sf]{subfig}
\usepackage{verbatim}
\def\BibTeX{{\rm B\kern-.05em{\sc i\kern-.025em b}\kern-.08em
    T\kern-.1667em\lower.7ex\hbox{E}\kern-.125emX}}
\usepackage{balance}
\usepackage{bm}
\usepackage{threeparttable}

\begin{document}

\title{Non-Diffracting Beams for Near-Field Millimeter-Wave Communications: Advantage Regimes Under Aperture and Blockage Constraints}

\author{Yifeng Qin,~\IEEEmembership{Member,~IEEE,}
        Jing Chen,
        Zhi Hao Jiang,~\IEEEmembership{Member,~IEEE,}
        Zhining Chen,~\IEEEmembership{Fellow,~IEEE,}\\
        and~Yongming Huang,~\IEEEmembership{Fellow,~IEEE,}
\thanks{This work was supported by the ... . (Corresponding author: \textit{Yifeng Qin}.)}
\thanks{Y. Qin and J. Chen are with the Peng Cheng Laboratory, Shenzhen, 518052, China (e-mails: ee06b147@gmail.com, chenj12@pcl.ac.cn).}
\thanks{Z. H. Jiang is with the State Key Laboratory of Millimeter Waves, School of Information Science and Engineering, Southeast University, Nanjing 210096, China (e-mail: zhihao.jiang@seu.edu.cn).}
\thanks{Z. N. Chen is with the Department of Electrical and Computer Engineering, National University of Singapore, Singapore 117583 (e-mail: eleczn@nus.edu.sg).}
\thanks{Y. Huang is with the National Mobile Communication Research Laboratory and the School of Information Science and Engineering, Southeast University, Nanjing 210096, China, and also with the Purple Mountain Laboratories, Nanjing 211111, China (e-mail: huangym@seu.edu.cn).}
}

\markboth{IEEE Transactions on Communications,~Vol.~XX, No.~X, 2026}%
{Qin \MakeLowercase{\textit{et al.}}: Aperture-Constrained Non-Diffracting Beams for Blockage-Resilient Near-Field Millimeter-Wave Communications}


\maketitle

\begin{abstract}
Near-field blockage changes the beam-design objective in millimeter-wave links: maximizing the unblocked on-axis gain does not necessarily maximize blocked-link performance. This paper studies when phase-only, aperture-constrained non-diffracting (ND) beams provide a blocked-link advantage over equal-aperture, equal-power conventional reference beams. We develop a unified annular-spectrum framework that generates isotropic Bessel-like and anisotropic Mathieu-like beams under discrete phased-array constraints, and a geometry-aware analysis centered on three propagation landmarks: the peak-intensity distance, the crossover distance, and an effective post-blockage recovery distance. Their relationship yields a recovery-before-crossover condition linking blockage size, depth, cone angle, and usable ND range, and motivates a blocked-link gain ratio that maps directly onto an achievable-rate gap at every operating SNR. The analysis also explains why anisotropic Mathieu-like beams can outperform isotropic ones under direction-dependent blockage. Monte Carlo simulations verify the predicted advantage regimes, an auxiliary comparison against a near-field focusing baseline confirms that the advantage persists against an unblocked-optimal array, and sensitivity studies over cone-angle choice and partial-transmission blockers show that the opaque-screen picture is a conservative reading of the underlying physics. The results identify Bessel-like and Mathieu-like beams as practical candidates for blockage-resilient near-field communications.
\end{abstract}


\begin{figure*}[!t]
    \centering
    \includegraphics[width=\textwidth]{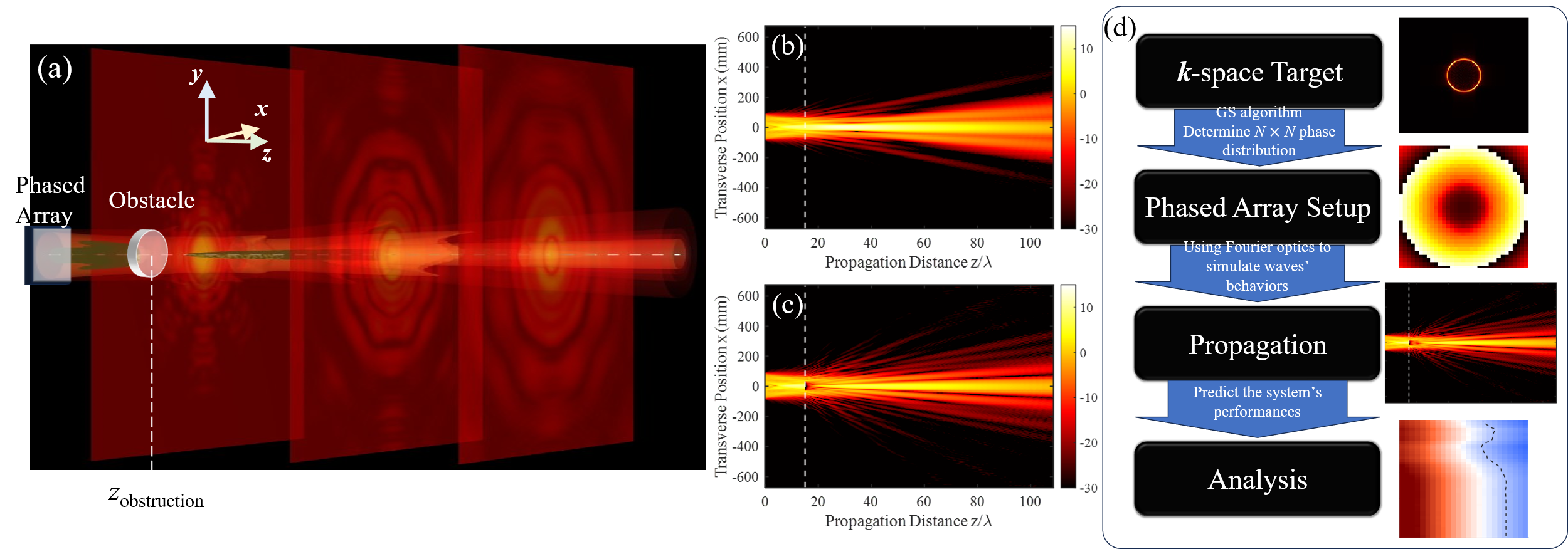}
    \caption{End-to-end framework for evaluating aperture-constrained ND beams in radiative near-field (RNF) blockage scenarios. (a)~3D visualization of ND-beam self-healing after obstruction, with transverse intensity profiles at several propagation distances. (b)~XZ-plane intensity map (dB) of the unblocked ND beam. (c)~XZ-plane intensity map of the blocked ND beam. (d)~Unified workflow: generalized $\bm{k}$-space target, phased-array synthesis via a holographic algorithm, angular-spectrum propagation, and link-level analysis.}
    \label{fig:framework}
\end{figure*}

\section{Introduction}
\label{sec:introduction}

\IEEEPARstart{A}{t} millimeter-wave (mmWave) and terahertz (THz) frequencies, high free-space path loss is compensated by packing a large number of small antenna elements into a physically large aperture, yielding high directional gain through electrically large arrays \cite{ref1,ref2,ref3}. A well-known consequence of this aperture scaling is that the Fraunhofer far-field boundary recedes, placing many intended user positions inside the radiative near field of the transmitter, where the received wavefront retains the spatial structure of the aperture rather than reducing to a simple angular pattern \cite{ref4,ref5,ref6,ref7,ref8}. This near-field operating regime is in fact an inherent feature of wideband, high-frequency communication links \cite{petrov_nf}. It enables spatially resolved transmission---near-field sensing, user-specific beam focusing, spatial multiplexing at sub-wavelength granularity---but at the same time makes the link vulnerable to near-field blockage. An obstacle situated between the aperture and the intended focal region can disrupt the wavefront structure and substantially degrade the received field, even when the unblocked beamforming gain is large \cite{ref9,ref10}.

The large number of phase-controllable elements in such arrays provides a high-dimensional aperture design space. Conventional beamforming uses this space to maximize focal-point gain, but the same degrees of freedom can also be used to shape the transmitted wavefront for intrinsic blockage tolerance---without requiring additional infrastructure such as relays or reconfigurable intelligent surfaces \cite{ref11,ref12,ref13,ref14}. Non-diffracting (ND) beams are one class of structured wavefronts that exploit this possibility. Since the ideal Bessel formulation of \cite{ref16}, ND beams with extended propagation and partial post-obstacle reconstruction (self-healing) have been studied extensively in optics and electromagnetics. Recent THz and sub-THz studies have applied these ideas to communication links through obstacle-tolerant Bessel transmission, analytical blockage modeling, and measurement-based investigations \cite{ref9,ref17,ref18,ref19,ref20,ref22,ref23}. Quasi-optical implementations using axicons, metalenses, and metasurfaces have further demonstrated extended-range Bessel generation \cite{ref24,ref25}, and recent phased-array studies of Airy/curving beams and closed-form Bessel designs have brought ND wavefront engineering onto practical array hardware \cite{droulias2025,uchimura2025}.

Despite these advances, translating the self-healing property into a communication-system link budget requires answering several questions that the existing literature has not addressed jointly: (i)~under identical aperture and radiated-power budgets, when does an ND beam outperform a conventional reference beam in a blocked near-field link? (ii)~what geometric quantities determine the size and location of the advantage regime? and (iii)~does the anisotropy of Mathieu-like beams provide a meaningful orientation-based robustness gain over isotropic Bessel-like beams against directionally structured blockers? All three questions must be answered under a common set of realistic hardware constraints---finite aperture, discrete phase quantization, and phase-only control---to produce conclusions that a system designer can act on.

This paper addresses these questions for finite-aperture, phase-only phased arrays at mmWave. The end-to-end methodology is illustrated in Fig.~\ref{fig:framework}. The main contributions are as follows:
\begin{enumerate}
    \item \textbf{Unified annular-spectrum synthesis:} We develop a common $\bm{k}$-space framework in which isotropic Bessel-like and anisotropic Mathieu-like beams emerge as two specializations of a single parameterized target spectrum, synthesized under the same aperture, amplitude taper, and phase-only constraints. This enables fair beam-to-beam comparisons under identical hardware and radiated-power budgets.

    \item \textbf{Geometry-aware blocked-link analysis:} We identify three propagation landmarks---the peak-intensity distance, the crossover distance relative to a boresight reference, and an effective post-blockage recovery distance---and derive a necessary geometric admissibility condition ($H>1$) for self-healing-driven advantage. An orientation-based degree of freedom for anisotropic spectra and a link-level beam-selection rule that maps directly to an achievable-rate gap complete the analytical framework.

    \item \textbf{Comprehensive numerical evaluation:} Through $73{,}920$-trial Monte Carlo simulation across eight realistic blocker families, an auxiliary near-field focusing comparison, a cone-angle robustness study, and a partial-transmission sensitivity analysis, we quantify the bounded advantage regimes and identify the specific geometric and spectral conditions under which ND beams are beneficial.
\end{enumerate}

The rest of this paper is organized as follows. Section~II reviews finite-aperture ND beams and introduces the unified spectral parameterization. Section~III presents the aperture-constrained synthesis method and the propagation/blockage model. Section~IV develops the geometry-aware blocked-link analysis and beam-selection criterion. Section~V describes the simulation setup and Monte Carlo protocol. Section~VI presents the numerical results. Section~VII discusses model validity, scope, and limitations, and Section~VIII concludes.

\section{Finite-Aperture ND Beams Under Array Constraints}

\subsection{Ideal Non-Diffracting Beams and Finite-Aperture Reality}
Ideal non-diffracting beams, such as the fundamental Bessel beam, are characterized by a transverse intensity profile that remains invariant as it propagates along the optical axis \cite{ref16}. Mathematically, their component plane waves share a common longitudinal wavenumber $k_z$, meaning their wave vectors $\bm{k} = (k_x, k_y, k_z)$ lie on a cone. The transverse wavenumbers thus form a perfect ring in $\bm{k}$-space, satisfying $k_x^2 + k_y^2 = k_{\rho}^2$, where $k_{\rho} = k_0 \sin\theta_c$ and $\theta_c$ is the beam's cone angle. 

While ideal ND beams require an infinite aperture and carry infinite energy, any practical realization is subject to finite-aperture truncation. When generated by a finite phased array, the beam's non-diffracting property is bounded to a finite axial range, accompanied by an inherent formation zone. Furthermore, practical arrays impose spatial discretization (element spacing) and component constraints (such as phase-only control). Our framework makes these implementation constraints explicit, treating them as necessary conditions rather than non-idealities, ensuring all candidate beams are evaluated under identical physical hardware limits.

\subsection{Unified Annular-Spectrum Parameterization}
To systematically evaluate structured beams, we define a generalized target spectrum in $\bm{k}$-space. The conical wave property translates to an annular ring in spatial frequency. We parameterize this continuous target spectrum with a Gaussian-profiled ring and an azimuthal angular modulation function:
\begin{equation}
|F_{\text{target}}(k_x, k_y)| \propto \exp\left(-\frac{(\tilde{\rho}_k-1)^2}{2\sigma_k^2}\right) \cdot B(\phi_k)
\label{eq:target_spectrum}
\end{equation}
where $\tilde{\rho}_k = \sqrt{k_x^2/k_a^2 + k_y^2/k_b^2}$ is the dimensionless elliptic radial coordinate controlling the ring ellipticity, $\sigma_k$ sets the spectral thickness, and $\phi_k = \operatorname{atan2}(k_y,k_x)\in[-\pi,\pi)$ is the quadrant-resolved azimuthal angle. The angular modulation term is $B(\phi_k) = 1 + \epsilon_2 \cos(2\phi_k) + \epsilon_4 \cos(4\phi_k)$, with $(\epsilon_2,\epsilon_4)$ restricted to the region $B(\phi_k)\ge 0$ so that the target magnitude in \eqref{eq:target_spectrum} remains non-negative. The principal axes $k_a,k_b$ are chosen so that the annulus peaks at $|k_t|\sim k_0\sin\theta_c$. 

This parameterization serves as a unified spectral target. The central radii $k_a$ and $k_b$ govern the cone angle and transverse scale, while $\sigma_k$ dictates the ND range and field purity.

\subsection{Bessel-Like and Mathieu-Like Beams as Specializations}
By adjusting the parameters in \eqref{eq:target_spectrum}, classic structured beams emerge as special cases of the unified framework:
\begin{itemize}
    \item \textbf{Isotropic Bessel-like beams:} setting $k_a = k_b$ and $\epsilon_2 = \epsilon_4 = 0$ yields a perfectly isotropic annular ring, producing a circularly symmetric Bessel beam geometry with geometry-agnostic blockage resilience (Fig.~\ref{fig:validation}(a)).
    \item \textbf{Anisotropic Mathieu-like beams:} setting $k_a \neq k_b$ elongates the annular ring into an ellipse, forming a Mathieu-like beam whose principal axis can be aligned with the dominant blocker orientation (Fig.~\ref{fig:validation}(b)).
\end{itemize}

The angular modulation $B(\phi_k)=1+\epsilon_2\cos(2\phi_k)+\epsilon_4\cos(4\phi_k)$ provides a second, orthogonal design lever: even on an isotropic annulus ($k_a=k_b$), a positive $\epsilon_2$ redistributes spectral mass along one azimuthal axis, producing a directionally weighted Bessel-like beam (Fig.~\ref{fig:validation}(c)) that trades geometry-agnostic robustness for increased robustness along the chosen axis. The higher harmonic $\epsilon_4$ enables finer angular shaping, including spectral notches that steer azimuthal nulls at prescribed $\phi_k$ directions (Fig.~\ref{fig:validation}(d)). All four variants in Fig.~\ref{fig:validation} are synthesized by the same GS algorithm (Section~III) under identical aperture, amplitude taper, and radiated-power constraints, so they can be compared fairly at the link-budget level without any reparameterization. In the analytical and Monte Carlo results that follow, we focus on the two endpoint regimes—pure isotropic Bessel-like and ellipticity-based Mathieu-like—because they isolate the two dominant levers; the angular-modulation variants are synthesized here only to establish that the unified framework covers them natively.

\section{Aperture-Constrained Synthesis and Propagation Model}

\subsection{Phase-Only Holographic Synthesis via GS Projection}
To convert the idealized continuous spectrum into an implementable phased-array pattern, we apply an iterative Gerchberg-Saxton (GS) holographic algorithm \cite{gerchberg1972}. The algorithm alternates between the physical array aperture plane and the $\bm{k}$-space, recursively applying the respective constraints. 

In the $\bm{k}$-space domain, the field magnitude is forced to match $|F_{\text{target}}|$ from \eqref{eq:target_spectrum}. Upon inverse-transforming to the aperture plane, the amplitude is constrained by the physical array window function $A_{\text{aperture}}$, forcing a phase-only solution. After sufficient iterations, the algorithm converges to a phase map $\Phi_{\text{array}} = \text{angle}(E_{\text{aperture}})$ that best approximates the target spectrum subject to the strict constraint of uniform-amplitude (or fixed-taper) radiation.

\begin{algorithm}[t]
\DontPrintSemicolon
\caption{Holographic Synthesis of Aperture Phase}
\label{alg:holographic_synthesis}
\SetKwProg{Procedure}{procedure}{}{end procedure}
\Procedure{\textsc{Generate\_ND\_Phase}(params)}{
    \textbf{Initialize:} $|F_{\text{target}}|$ from (\ref{eq:target_spectrum}); aperture amplitude constraint $A_{\text{aperture}}$; random phase $E_{\text{aperture}}$\;
    $F_{\text{current}} \leftarrow \text{FFT}(E_{\text{aperture}})$.\;
    \For{$i = 1$ \KwTo $N_{\text{iterations}}$}{
        $F_{\text{current}} \leftarrow |F_{\text{target}}| \cdot \exp(j \cdot \text{angle}(F_{\text{current}}))$\;
        $E_{\text{aperture}} \leftarrow \text{IFFT}(F_{\text{current}})$\;
        $E_{\text{aperture}} \leftarrow A_{\text{aperture}} \cdot \exp(j \cdot \text{angle}(E_{\text{aperture}}))$\;
        $F_{\text{current}} \leftarrow \text{FFT}(E_{\text{aperture}})$\;
    }
    \Return $\text{angle}(E_{\text{aperture}})$\;
}
\end{algorithm}

Each GS iteration consists of two 2D FFTs on a $N\times N$ grid and two pointwise projections, for a per-iteration cost of $\mathcal{O}(N^2\log N)$. With $N=64$ and $N_{\text{iter}}=200$ iterations, a single phase map requires on the order of $10^5$ complex operations and synthesizes in milliseconds on a conventional CPU, so the phase profile can be precomputed offline for each $(\theta_c, k_a/k_b, \epsilon_2)$ configuration and selected at run time from a compact codebook. Real-time adaptation to slow-moving blockers therefore reduces to a codebook lookup rather than a live GS run.

\subsection{Fairness Baseline and Fixed Parameters}
The complete set of array, synthesis, and propagation parameters is collected in Table~\ref{tab:params}. All candidate beams and the reference beam share the same $64\times 64$ aperture, the same Super-Gaussian amplitude taper, the same total radiated power, the same ASM propagation operator, the same thin-screen blockage model, and the same receive combining rule; the only difference is the aperture phase pattern. The reference beam---referred to as the boresight reference---uses a uniform (zero) aperture phase and therefore radiates without near-field re-phasing, isolating the contribution of the ND phase structure. A separate auxiliary comparison against a near-field focusing baseline, whose aperture phase is optimized to maximize the unblocked on-axis gain at the user coordinate, is reported in Section~VI-D.

\subsection{Propagation and Blockage Model}
Continuous wave propagation from the aperture ($z=0$) to the receiver ($z = z_{\text{eval}}$) is computed scalarly using the angular spectrum method (ASM) representing the Rayleigh-Sommerfeld integral. Assuming the $\exp(j\omega t)$ time convention, the transverse field is advanced via the free-space transfer function:
\begin{equation}
H(k_x, k_y, \Delta z) = \exp\left(-j \Delta z \sqrt{k_0^2 - k_x^2 - k_y^2}\right).
\label{eq:transfer_func}
\end{equation}

Obstacles are modeled as perfectly opaque thin screens placed normally to the propagation axis at depth $z_{\text{obs}}$. The post-blockage field is the element-wise product of the incident field and the binary obstacle mask $M(x, y)$:
\begin{equation}
E(x, y, z_{\text{obs}}^+) = E(x, y, z_{\text{obs}}^-) \cdot M(x, y).
\label{eq:obstacle}
\end{equation}
This framework captures rigorous near-field diffraction, including both the self-healing interference of the ND beam and knife-edge (Poisson-Arago) diffraction.

The full set of fixed array, synthesis, and propagation parameters used throughout Sections~V and~VI is collected in Table~\ref{tab:params}.

\begin{table}[!t]
\centering
\caption{Fixed Array, Synthesis, and Propagation Parameters}
\label{tab:params}
\renewcommand{\arraystretch}{1.15}
\begin{threeparttable}
\begin{tabular}{@{}l l@{}}
\hline
\textbf{Parameter} & \textbf{Value} \\
\hline
Transmit aperture & $64\times 64$ uniform planar array \\
Element spacing & $\lambda/2$ \\
Effective aperture & $32\lambda\times 32\lambda$ \\
Phase quantization & 6 bits \\
Amplitude taper & Super-Gaussian (5-bit quantized) \\
Target ND cone half-angle $\theta_c$ & $\approx 7^\circ$ \\
GS iterations & $200$ \\
Propagation operator & non-paraxial ASM \eqref{eq:transfer_func} \\
Obstacle model & opaque thin screen \eqref{eq:obstacle} \\
Receive array & $2\times 2$ UPA, $0.49\lambda$ spacing \\
Receive combining & element-wise phase-conjugate \\
Blocker library size & 8 named families (Section V-B) \\
Size-scale sweep per family & $\text{sf}\in[0.70,\,1.35]$, $21$ points \\
Random placements per (family, sf) & $40$ \\
Evaluation depths per placement & $11$ in $[z_{\text{peak}},z_{\text{crossover}}]$ \\
Monte Carlo trial budget & $73{,}920$ trials total \\
Random seed & $20240901$ (reused across arms) \\
\hline
\end{tabular}
\end{threeparttable}
\end{table}

\section{Geometry-Aware Advantage Analysis}

Self-healing describes a restoration of the field pattern, but it does not by itself guarantee a communication advantage over a conventional reference beam radiating from the same aperture. This section develops a geometry-aware analysis that translates the field-pattern property into a link-level condition, in terms of the blockage dimensions, the user position, and the structural parameters of the ND beam.

\subsection{Propagation Landmarks: $z_{\text{peak}}$ and $z_{\text{crossover}}$}
Let $I_{\text{ND}}(z)\triangleq|E_{\text{ND}}(0,0;z)|^2$ and $I_{\text{REF}}(z)\triangleq|E_{\text{REF}}(0,0;z)|^2$ denote the unblocked on-axis intensities of the ND beam and the reference beam under identical aperture, radiated power, and propagation operator. We define the two link-level landmarks
\begin{align}
z_{\text{peak}} &\triangleq \arg\max_{z>0}\, I_{\text{ND}}(z), \\
z_{\text{crossover}} &\triangleq \inf\{z>z_{\text{peak}}:\, I_{\text{ND}}(z)=I_{\text{REF}}(z)\}.
\end{align}
In the designs evaluated in this paper, $I_{\text{ND}}(z)$ peaks once in the radiating near field and then decays below $I_{\text{REF}}(z)$ at a unique crossing, so $z_{\text{crossover}}$ is well defined and the interval $[z_{\text{peak}},\,z_{\text{crossover}})$ is the unblocked-link advantage window of the ND beam: outside it, the reference beam already dominates on axis, and no blockage-driven mechanism can restore ND superiority. Every blocked-link statement in the remainder of this section refers to an evaluation distance $z_{\text{eval}}\in[z_{\text{peak}},z_{\text{crossover}})$.

\begin{figure*}[!t]
    \centering
    \includegraphics[width=\textwidth]{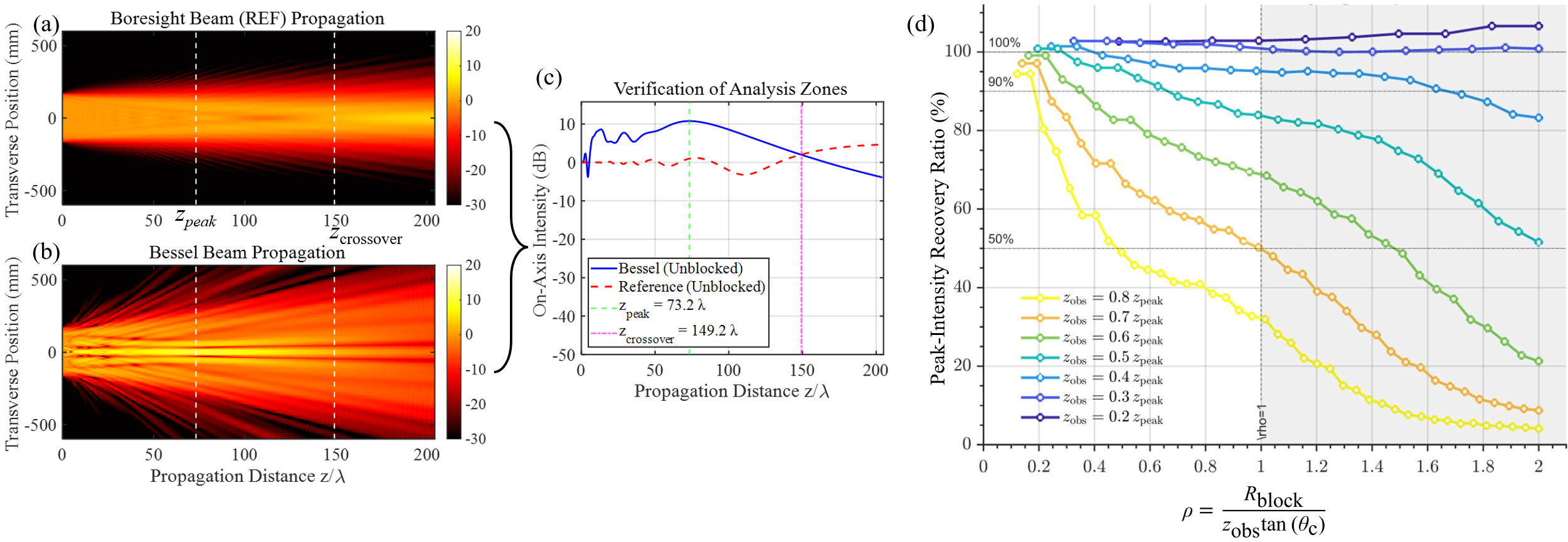}
    \caption{(a) Boresight reference propagation and (b) Bessel-like ND propagation, shown as XZ-plane max-intensity maps (dB). White dashed lines mark $z_{\text{peak}}$ and $z_{\text{crossover}}$. (c) Unblocked on-axis intensity versus distance, verifying $z_{\text{peak}}$ (green) and $z_{\text{crossover}}$ (magenta). (d) Peak-intensity recovery ratio versus $\rho=R_{\text{block}}/(z_{\text{obs}}\tan\theta_c)$ for centered circular blockers at various depths $z_{\text{obs}}/z_{\text{peak}}$. A mild overshoot above $100\%$ at small $\rho$ arises from constructive Fresnel edge diffraction.}
    \label{fig:metrics}
\end{figure*}

\subsection{Recovery Geometry and Admissibility Factor}
We now derive a necessary condition under which the self-healing mechanism can deliver an on-axis blocked-link advantage at the user location. The condition reduces to a single geometric inequality.

Consider a blocker of effective radius $R_{\text{eff}}$ placed at depth $z_{\text{obs}}$. The conical angular components of the ND beam, propagating at angle $\theta_c$ to the optical axis, re-intersect the axis after a geometric healing distance
\begin{equation}
z_{\min} \approx \frac{R_{\text{eff}}}{\tan\theta_c},
\label{eq:zmin}
\end{equation}
since the outermost surviving components must travel at least $R_{\text{eff}}$ transversely to reach the axis. Self-healing therefore delivers a usable on-axis field only if the reconstruction completes within the unblocked advantage window of Section~IV-A, i.e., before the reference beam dominates on axis. Writing $z_{\mathrm{w}}\triangleq\min\{z_{\text{eval}},z_{\text{crossover}}\}$ for the available downstream budget, the reconstruction condition is $z_{\text{obs}}+z_{\min}<z_{\mathrm{w}}$, or equivalently
\begin{equation}
H \;\triangleq\; \frac{z_{\mathrm{w}} - z_{\text{obs}}}{z_{\min}}
  \;=\; \frac{(z_{\mathrm{w}} - z_{\text{obs}})\tan\theta_c}{R_{\text{eff}}} \;>\; 1.
\label{eq:Hfactor}
\end{equation}
The dimensionless factor $H$ is the ratio of the downstream budget available to the beam to the healing distance required by the geometry. When $H>1$, the available downstream distance is sufficient for the beam to reconstruct on axis; when $H\le 1$, no annular-spectrum design can recover the missing downstream distance. This provides the first operating condition of the framework.

\textbf{Proposition 1 (Necessary geometric admissibility).} \emph{Let $z_{\text{eval}}\in[z_{\text{peak}},z_{\text{crossover}})$ and $z_{\mathrm{w}}=\min\{z_{\text{eval}},z_{\text{crossover}}\}$. A self-healing-driven ND advantage over the reference beam at $z_{\text{eval}}$ requires}
\begin{equation}
H \;>\; 1.
\end{equation}

\begin{IEEEproof}
We give a geometric proof sketch. Let $E_{\text{ND}}(0,0;z)$ denote the on-axis blocked-link field of the ND beam. The self-healing mechanism that produces this field is the re-interference of the conical plane-wave components that were not intercepted by the obstacle; in the paraxial limit at cone half-angle $\theta_c$, those components re-intersect the optical axis at a downstream distance of at least $R_{\text{eff}}/\tan\theta_c=z_{\min}$ past the obstacle plane. For all $z\in[z_{\text{obs}}, z_{\text{obs}}+z_{\min})$ the self-healing contribution to $|E_{\text{ND}}(0,0;z)|$ is therefore absent, and any nonzero on-axis response inside this interval is dominated by edge-diffraction terms that are not part of the self-healing mechanism analyzed here. If $z_{\text{obs}}+z_{\min}\ge z_{\mathrm{w}}$, this no-self-healing interval covers the entire evaluation window $[z_{\text{obs}},z_{\mathrm{w}})$, and no self-healing-driven advantage can be realized at $z_{\text{eval}}$. The condition $z_{\text{obs}}+z_{\min}<z_{\mathrm{w}}$ rearranges to $(z_{\mathrm{w}}-z_{\text{obs}})/z_{\min}>1$, i.e., $H>1$.
\end{IEEEproof}

\subsection{Directional Robustness of Anisotropic Spectra}
Proposition~1 provides a necessary longitudinal condition but, at fixed $(\theta_c,R_{\text{eff}},z_{\text{obs}})$, does not depend on the azimuthal structure of the ND spectrum. The anisotropy of the Mathieu-like family introduces an additional degree of freedom by redistributing the angular content of the surviving spectral mass. Obtaining a closed-form bound for this directional effect is nontrivial: the on-axis blocked field is a \emph{coherent} sum of surviving plane-wave components, and its magnitude depends not only on the proportion of spectral mass that is not intercepted by the obstacle but also on the phase alignment of those components at $z_{\text{eval}}$. This phase alignment is sensitive to the discrete finite-aperture GS synthesis and to the obstacle boundary geometry, so a strict pointwise inequality is not available. We therefore formulate the orientation mechanism as an observation based on a coherent-sum proxy. The proxy captures the monotonicity of the mechanism under rotation of the spectrum and is verified numerically in Section~VI-C.

Let $\Phi_{\text{blk}}$ denote the angular sector of the transverse $\bm{k}$-plane whose corresponding plane-wave components are intercepted by the blocker, and let $S(F)\triangleq\iint|F(k_x,k_y)|\,dk_x\,dk_y$. In the ND design regime, the surviving plane-wave components are engineered to add in phase on axis, so the on-axis blocked field magnitude is upper-bounded---and approximately attained---by the $L_1$-norm of the surviving spectrum. We therefore define the \emph{surviving coherent-sum proxy} $\xi_{\text{surv}}\in(0,1]$ as
\begin{equation}
\xi_{\text{surv}} \;\triangleq\; \frac{S_{\text{surv}}}{S(F_{\text{target}})},
\quad
S_{\text{surv}} \triangleq \!\!\iint\limits_{\phi_k\notin\Phi_{\text{blk}}}\!\!|F_{\text{target}}(k_x,k_y)|\,dk_x dk_y.
\label{eq:xisurv}
\end{equation}
For an isotropic Bessel-like target ($k_a=k_b$, $\epsilon_2=\epsilon_4=0$), $\xi_{\text{surv}}$ equals the fractional angular coverage of the unblocked sector and is therefore independent of beam orientation. For an anisotropic Mathieu-like target ($k_a\neq k_b$ or $\epsilon_2\neq 0$), $\xi_{\text{surv}}$ depends on the orientation of the spectral ring relative to $\Phi_{\text{blk}}$ and is maximized when the high-density lobes are rotated out of $\Phi_{\text{blk}}$. (Note that for anisotropic spectra, $\theta_c$ is defined along the receiver-facing principal axis).

\textbf{Observation 2 (Orientation degree of freedom).} \emph{For equal-aperture, equal-power ND beams evaluated at a common downstream budget $(\theta_c,R_{\text{eff}},z_{\text{obs}},z_{\text{eval}})$, the blocked-link on-axis coherent sum is governed empirically by the surviving coherent-sum proxy $\xi_{\text{surv}}$. Rotating an anisotropic spectrum so that its high-density lobes lie outside $\Phi_{\text{blk}}$ strictly increases $\xi_{\text{surv}}$ and provides a geometry-adaptive orientation degree of freedom that an isotropic Bessel-like beam does not possess.}

\emph{Sketch.} Under the common angular-spectrum propagator, the on-axis blocked field at a distance $\Delta z$ past the obstacle is the phasor coherent sum of the surviving plane-wave components, whose magnitude obeys the triangle inequality $|E(0,0;\Delta z)|\le(2\pi)^{-2}\iint|F_{\text{surv}}|\,dk_x\,dk_y$. Rotating an anisotropic target spectrum to concentrate amplitude outside $\Phi_{\text{blk}}$ is a pure geometric action on the aperture-plane field: it leaves $R_{\text{eff}}$, $\theta_c$, and the downstream budget unchanged, and strictly increases $\xi_{\text{surv}}$. The $L_1$ quantity is a \emph{coherent-sum proxy} rather than a tight bound---synthesized ND beams on finite apertures recover a geometry-dependent fraction of this upper bound because of sidelobe spread and sub-optimal phase alignment---but $\xi_{\text{surv}}$ is orientation-monotone by construction, and the empirical per-shape data in Section~VI-C confirm that the orientation-induced change in $\xi_{\text{surv}}$ moves together with the orientation-induced change in blocked-link coherent sum across the blocker families. \hfill$\square$

\subsection{Complete Decision Rule and Formal Performance Metrics}
Proposition~1 and Observation~2 act on two orthogonal levers---geometric headroom ($H$) and spectral orientation ($\xi_{\text{surv}}$)---and neither alone determines the link outcome. The actual decision is made at the link-budget level, which requires reducing the continuous scalar field of Section~III to a discrete transmit--receive channel. Let $\{\mathbf{r}_n^{(\text{T})}\}_{n=1}^{N}$ and $\{\mathbf{r}_m^{(\text{R})}\}_{m=1}^{M}$ be the positions of the transmit and receive antenna elements, modeled as isotropic point sources and samplers; the $(m,n)$-th entry of the blocked near-field MIMO channel matrix $\mathbf{H}_{\text{blk}}(z_{\text{eval}})\in\mathbb{C}^{M\times N}$ is the scalar ASM response from $\mathbf{r}_n^{(\text{T})}$ through the obstacle mask $M(x,y)$ at $z_{\text{obs}}$ to $\mathbf{r}_m^{(\text{R})}$ at $z_{\text{eval}}$, computed under \eqref{eq:transfer_func}--\eqref{eq:obstacle}. With this identification, the normalized aperture-plane precoder $\mathbf{w}_b\in\mathbb{C}^{N\times 1}$ absorbs the amplitude taper and the synthesized phase pattern for beam $b\in\{\text{ND},\text{REF}\}$, and the receive-side field vector at the user array is $\mathbf{e}_b = \mathbf{H}_{\text{blk}}\mathbf{w}_b$. Under equal total radiated power and identical phase-conjugate combining $\mathbf{f}_b = \exp(j\angle\mathbf{e}_b)$, the blocked-link advantage ratio is
\begin{equation}
\Gamma_{\text{adv}}(z_{\text{eval}})
\triangleq
\frac{|\mathbf{f}_{\text{ND}}^H \mathbf{H}_{\text{blk}}\mathbf{w}_{\text{ND}}|^2}
     {|\mathbf{f}_{\text{REF}}^H \mathbf{H}_{\text{blk}}\mathbf{w}_{\text{REF}}|^2},
\qquad
\Delta P_{\text{rx}}\,(\text{dB})\triangleq 10\log_{10}\Gamma_{\text{adv}}.
\label{eq:delta_snr}
\end{equation}

\textbf{Decision Rule (Beam selection under blockage).} At $z_{\text{eval}}\in[z_{\text{peak}},z_{\text{crossover}})$, select the ND beam if $\Gamma_{\text{adv}}(z_{\text{eval}})>1$, and select the reference beam otherwise.

Proposition~1 and Observation~2 play complementary roles in this rule. Proposition~1 operates as a zero-cost geometric pre-screen: whenever $H\le 1$, the rule can return the reference beam without computing any field, because the necessary condition for self-healing-driven advantage has already failed. Observation~2 identifies the orientation of the anisotropic spectrum that maximizes $\xi_{\text{surv}}$, and hence the attainable $\Gamma_{\text{adv}}$, on each admissible geometry. The rule applies pairwise to any two candidate beams $b_1,b_2$ synthesized under the shared aperture, power, and propagation constraints: $b_1$ is preferred if $|\mathbf{f}_{b_1}^{H}\mathbf{H}_{\text{blk}}\mathbf{w}_{b_1}|^2>|\mathbf{f}_{b_2}^{H}\mathbf{H}_{\text{blk}}\mathbf{w}_{b_2}|^2$; in particular, the comparison between an isotropic Bessel-like and an anisotropic Mathieu-like candidate in Section~VI-C is this pairwise form with the reference arm replaced by the competing ND arm.

This three-step chain---$(z_{\text{peak}},z_{\text{crossover}})\rightarrow H\rightarrow\xi_{\text{surv}}\rightarrow\Gamma_{\text{adv}}$---is the analytical spine used for all Monte Carlo evaluations in Sections~V--VI.

\begin{figure}[!t]
    \centering
    \includegraphics[width=\columnwidth]{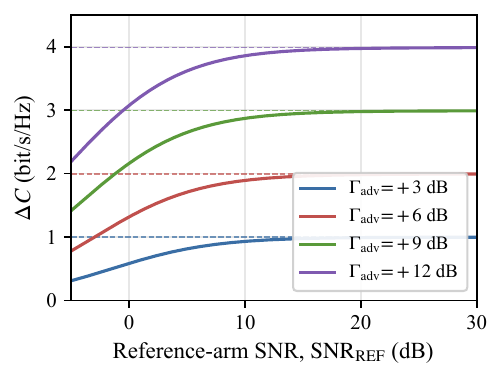}
    \caption{Rate gap $\Delta C$ between the ND and reference arms, as a function of the reference-arm blocked-link SNR, for four representative values of the advantage ratio $\Gamma_{\text{adv}}=\{+3,+6,+9,+12\}$~dB. Dashed lines indicate the high-SNR asymptotes $\log_2\Gamma_{\text{adv}}$. At moderate-to-high SNR, the dB-scale metric $\Delta P_{\text{rx}}$ reported throughout Section~VI is a near-linear surrogate for the achievable-rate gap.}
    \label{fig:rate_map}
\end{figure}

$\Delta P_{\text{rx}}$ is a power ratio, but the quantity that a communication system cares about is an achievable rate. The two are linked directly. Writing $P_b\triangleq|y_b|^2$ for the combiner output power of beam $b\in\{\text{ND},\text{REF}\}$ and $N_0$ for the noise power, the per-arm single-stream achievable rate is $C_b = \log_2(1+P_b/N_0)$, and the rate gap at $z_{\text{eval}}$ follows as
\begin{equation}
\Delta C \;=\; \log_2\!\left(\frac{1+\text{SNR}_{\text{REF}}\,\Gamma_{\text{adv}}}{1+\text{SNR}_{\text{REF}}}\right),
\label{eq:delta_C}
\end{equation}
where $\text{SNR}_{\text{REF}}\triangleq P_{\text{REF}}/N_0$ is the blocked-link SNR seen by the reference arm. The mapping has two regimes: at high SNR, $\Delta C\to\log_2\Gamma_{\text{adv}}$, so a $+3$~dB value of $\Delta P_{\text{rx}}$ corresponds to one bit/s/Hz; at low SNR, the additive unity in $\log_2(1+\cdot)$ compresses the gap. Fig.~\ref{fig:rate_map} shows both regimes for four representative values of $\Gamma_{\text{adv}}$. The beam-selection rule stated above therefore coincides with the rate-maximizing rule at every $\text{SNR}_{\text{REF}}$, and the advantage regime maps of Section~VI can be read directly as rate maps once a reference operating point is fixed. For clarity, we evaluate $\Delta C$ as a single-stream gap through the dominant spatial mode of the receive array; this isolates the contribution of the ND phase structure to the link budget from any spatial-multiplexing gain, which depends on additional factors (eigenmode geometry, channel estimation, stream coordination) and would confound the comparison at the level of an individual beam design.

\section{Simulation Setup and Monte Carlo Protocol}
This section specifies the receiver combining, the scenario library, and the trial budget used throughout Section~VI.

\subsection{Receiver and Combining Model}
The user equipment (UE) is a $2 \times 2$ uniform planar array (UPA) with $0.49\lambda$ element spacing implementing phase-conjugate combining. For an incident field $E$ sampled at the four element positions $(x_m,y_m)$, the receiver forms
\begin{equation}
y = \sum_{m=1}^{4} E(x_m, y_m, z_{\text{eval}}) \, e^{-j\arg E(x_m, y_m, z_{\text{eval}})}.
\end{equation}
The differential log-ratio $\Delta P_{\text{rx}}$ of \eqref{eq:delta_snr} is computed per trial under identical noise figures and is the received-power ratio between the ND arm and the reference arm. Phase-conjugate combining is adopted to align the receiver metric with the field-intensity comparison of Section~IV-D, not to model an optimal receiver. A different linear combiner would in general depend on the arm-specific receive field and can therefore modify both the absolute received powers and the relative margin between the two arms; a systematic study of receiver-design dependence is outside the scope of this paper.

\subsection{Scenario Library, Difficulty Coordinates, and Trial Budget}
Each Monte Carlo trial places a single thin-screen blocker between the transmit aperture and the user plane. Every trial is described by two normalized coordinates: a difficulty index $\rho$ that compresses the blocker geometry into a single number, and a normalized depth $t$ that compresses the user position relative to the ND landmarks. The difficulty index is
\begin{equation}
\rho \triangleq \frac{R_{\text{eff}}}{z_{\text{obs}} \tan\theta_c} \approx \frac{z_{\min}}{z_{\text{obs}}},
\label{eq:rho_difficulty}
\end{equation}
in which $R_{\text{eff}}$ aggregates the blocker's intrinsic half-size $R_{\text{block}}$ and the lateral offset $d$ between the blocker centroid and the optical axis; for centered circular blockers ($d=0$) such as those used in the admissibility sweep of Section~VI-A, $R_{\text{eff}}$ reduces to the geometric radius $R_{\text{block}}$, and for non-circular or off-axis blockers it is the smallest enclosing radius of the blocker's projection onto the cone-facing half-plane. We emphasize that $\rho$ is a one-dimensional \emph{difficulty coordinate} used to organize the Monte Carlo ensemble, and not a direct surrogate for the admissibility factor $H$ of Proposition~1: $\rho<1$ asserts only that the recovery distance $z_{\min}$ is smaller than the obstacle depth $z_{\text{obs}}$, whereas $H>1$ additionally requires that $z_{\min}$ be smaller than the downstream budget $z_{\mathrm{w}}-z_{\text{obs}}$. Large $\rho$ therefore corresponds to the high-difficulty end of the Monte Carlo coordinate, while the formal admissibility condition of Proposition~1 remains governed by $H$ and depends jointly on $z_{\text{eval}}$ and $z_{\text{crossover}}$. The normalized evaluation depth
\begin{equation}
t \triangleq \frac{z_{\text{eval}} - z_{\text{peak}}}{z_{\text{crossover}} - z_{\text{peak}}}\in[0,1]
\label{eq:t_norm}
\end{equation}
collapses different ND configurations onto a common axis; $t=0$ is the ND peak-intensity plane and $t=1$ is the unblocked-link crossover with the reference beam.

The blocker library contains eight named families chosen to span the realistic mmWave near-field obstruction set: \textbf{ArmBar}, \textbf{ChairBack}, \textbf{HumanSide}, \textbf{HumanTorso}, \textbf{PillarLarge}, \textbf{PillarSmall}, \textbf{TableEdge}, and \textbf{WallEdge}. Each family is realized through one of four geometric primitives (rectangular bar, circular disk, half-plane, full rectangle) with shape-specific aspect ratios. HumanTorso and HumanSide are kept strictly vertical; the remaining families receive uniform random in-plane rotations $\phi\sim\mathcal{U}[0,2\pi)$. A size-scale factor $\text{sf}\in[0.70,1.35]$ sweeps each family across $21$ scales so that the same geometry visits both well-inside the admissible regime ($\rho\ll 1$) and the saturation boundary ($\rho\approx 1.5$). The eight families together populate the $(\rho,t)$ plane densely rather than at isolated grid points, which is what makes the advantage maps in Section~VI interpretable as continuous regimes rather than as scenario-specific anecdotes.

For each (family, $\text{sf}$) pair, the simulator draws $40$ independent random placements—each with an independent in-plane offset $d$ and (for non-vertical families) an independent rotation $\phi$—and evaluates each placement at $11$ depths $z_{\text{eval}}$ uniformly spaced in $[z_{\text{peak}},z_{\text{crossover}}]$. The per-pair budget is therefore $11\times 40 = 440$ trials, and the overall budget is $8\times 21\times 11\times 40 = 73{,}920$ trials. The same random seed and the same $(\rho,t)$ realization are reused for every beam configuration compared in Section~VI, so that all reported win-rate differences come from the beam structure rather than from independent Monte Carlo noise.

\section{Results: Theory Verification and Advantage Regime Mapping}
\begin{figure}[!t]
    \centering
    \includegraphics[width=\columnwidth]{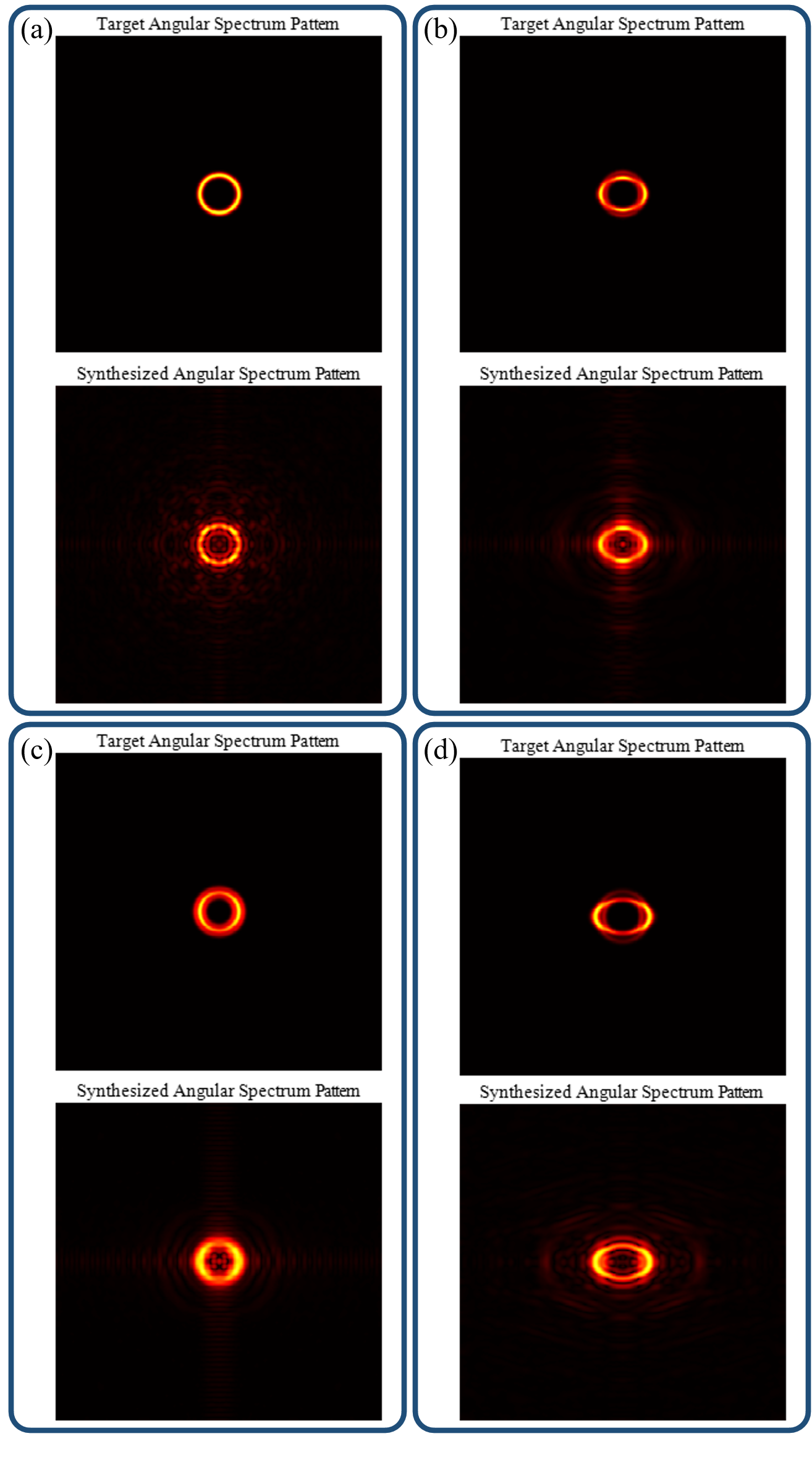}
    \caption{Validation of the unified $\bm{k}$-space synthesis framework. Each panel compares the ideal target spectrum (top) with the spectrum synthesized by the GS algorithm under the $64\times 64$ phased-array constraint (bottom). (a) Isotropic Bessel-like target ($k_a=k_b$, $\epsilon_2=\epsilon_4=0$). (b) Anisotropic Mathieu-like target ($k_a\neq k_b$, $\epsilon_2>0$). (c) Bessel-like target with a non-uniform angular power distribution. (d) Spectrally notched Bessel-like target designed to steer nulls in specific azimuthal directions.}
    \label{fig:validation}
\end{figure}

This section verifies the propositions developed in Section~IV using the Monte Carlo ensemble of Section~V. As a prerequisite, Fig.~\ref{fig:validation} confirms that the unified annular-spectrum framework synthesizes both isotropic Bessel-like and anisotropic Mathieu-like targets faithfully under the $64\times 64$ phased-array constraint (6-bit phase, 5-bit amplitude); the remainder of the section analyzes the blocked-link behavior of the synthesized beams.

\subsection{Verification of Geometric Admissibility ($H > 1$)}
Before evaluating the random-blockage ensemble, we first verify the geometric admissibility condition ($H > 1$) under a controlled baseline: a single centered circular blocker where the geometry is deterministic and absent of orientation-induced statistical variations. As a prerequisite, Fig.~\ref{fig:metrics}(a)--(c) identify the unblocked propagation landmarks $z_{\text{peak}}$ and $z_{\text{crossover}}$ for both beams, establishing the boundaries of the RNF advantage zone. 

Fig.~\ref{fig:metrics}(d) evaluates the recovery-distance mechanism by plotting the peak-intensity recovery ratio against the normalized obstacle radius $\rho$, stratified by the relative obstacle depth $z_{\text{obs}}/z_{\text{peak}}$. For shallow blockages (e.g., $z_{\text{obs}}/z_{\text{peak}}=0.2$), the downstream budget is large enough that $H>1$ holds even at moderately large $\rho$, and the beam reconstructs its on-axis intensity. As the obstacle is moved deeper into the field, the available post-blockage distance $z_{\text{crossover}}-z_{\text{obs}}$ decreases. The $\rho$ value at which the recovery curve drops sharply corresponds to $H$ crossing unity, which is consistent with Proposition~1: when the required healing distance exceeds the available downstream window, self-healing cannot complete within the usable ND range and the advantage collapses.

\begin{figure*}[!t]
    \centering
    \includegraphics[width=\textwidth]{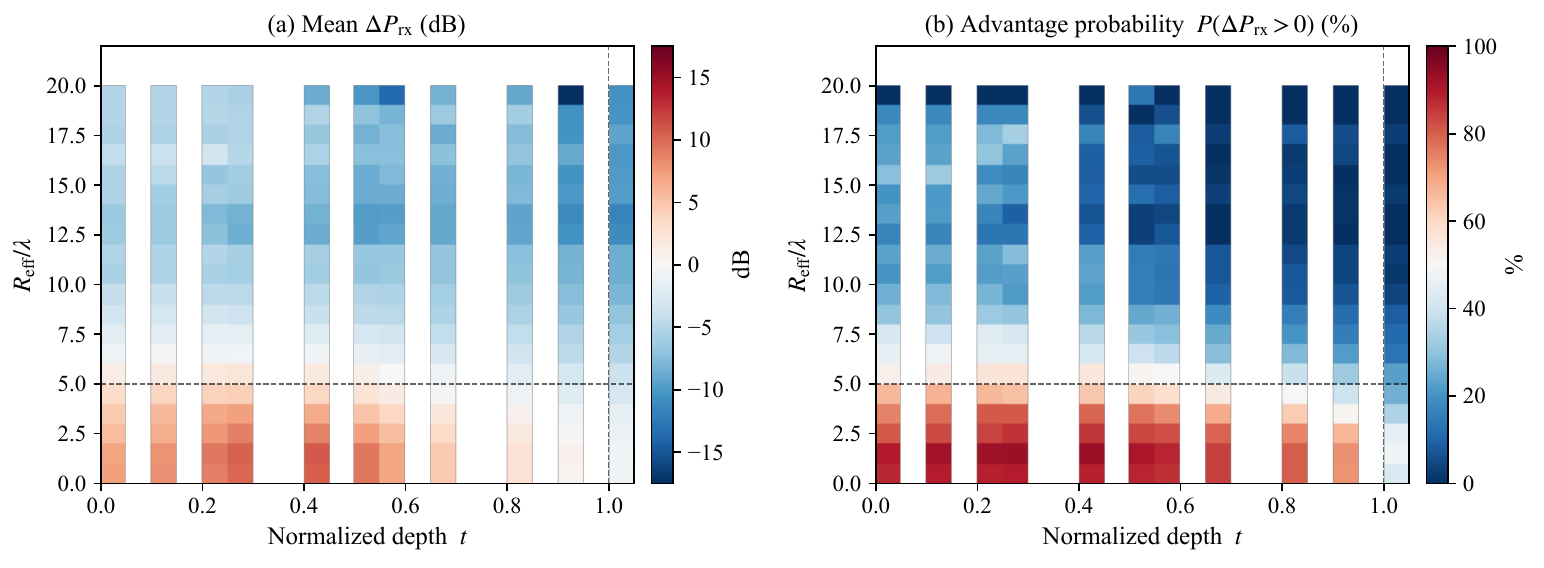}
    \caption{Isotropic Bessel-like beam versus the boresight reference over the seven finite-extent blocker families ($64{,}680$ trials). (a) Mean differential gain $\Delta P_{\text{rx}}$ (dB) on the $(R_{\text{eff}}/\lambda,\,t)$ plane. (b) Advantage probability $P(\Delta P_{\text{rx}}>0)$ (\%) on the same plane. The WallEdge family is excluded from this map because it is modeled as an infinite half-plane ($R_{\text{eff}}=\infty$); its depth-only statistics appear in Fig.~\ref{fig:stats}.}
    \label{fig:macro_map}
\end{figure*}

\begin{figure*}[!t]
    \centering
    \includegraphics[width=\textwidth]{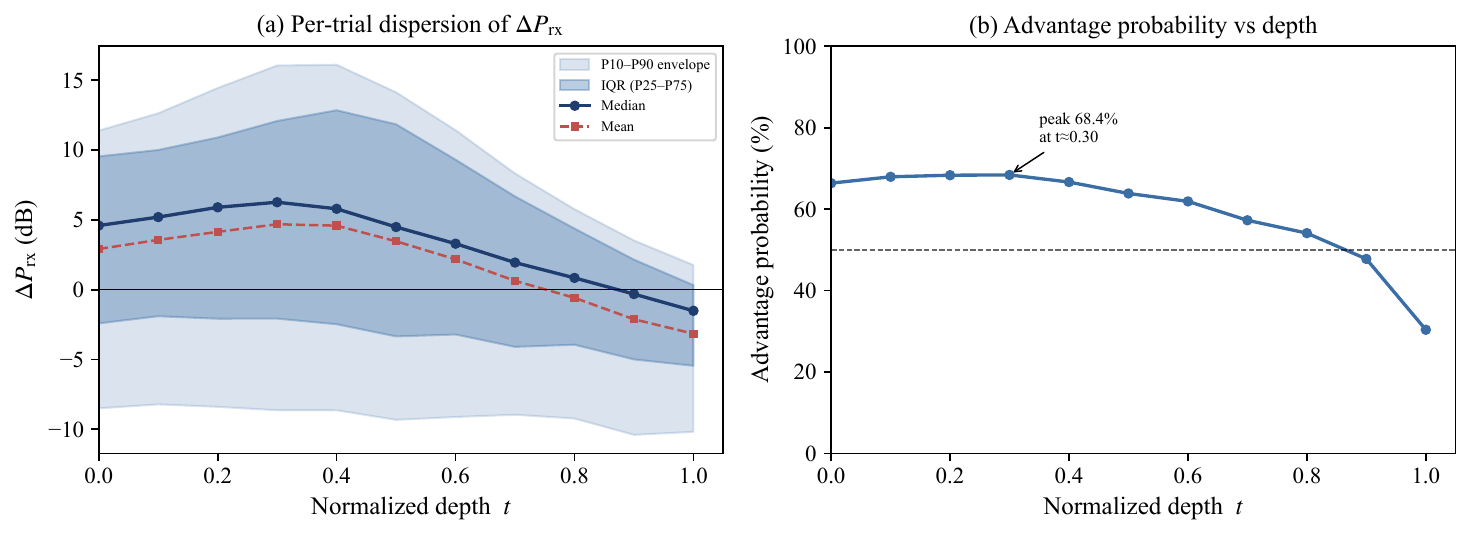}
    \caption{Per-trial dispersion of $\Delta P_{\text{rx}}$ for the isotropic Bessel-like beam against the boresight reference over the full eight-family ensemble ($73{,}920$ trials). (a) Mean, median, interquartile range, and $P_{10}$--$P_{90}$ envelope as a function of $t$. (b) Advantage probability versus $t$; the peak win rate is annotated.}
    \label{fig:stats}
\end{figure*}

\subsection{Bessel-Like Regime Maps}
Having confirmed that the admissibility factor $H$ bounds the advantage under deterministic centered blockage, we now evaluate the blocked-link performance across the full Monte Carlo ensemble, which spans multiple blocker sizes, shape families, and random lateral offsets. Fig.~\ref{fig:macro_map} reports the macro-averaged differential gain $\Delta P_{\text{rx}}$ of an isotropic Bessel-like beam against the boresight reference, organized along the two normalized geometric coordinates $(R_{\text{eff}}/\lambda,\,t)$. Of the eight blocker families introduced in Section~V-B, the seven finite-extent geometries with a well-defined effective radius $R_{\text{eff}}$ contribute the $7\times 21\times 11\times 40 = 64{,}680$ trials displayed in the map. The eighth family, WallEdge, is modeled as an infinite half-plane and therefore has $R_{\text{eff}}=\infty$ by construction; its $9{,}240$ trials cannot be placed on the $(R_{\text{eff}}/\lambda,\,t)$ axes and are instead summarized in the depth-only statistics of Fig.~\ref{fig:stats} and in the per-shape comparison of Section~VI-C. Two stratifications of the displayed $64{,}680$-trial subset make the predicted bounded-advantage structure quantitative.

Stratifying by obstacle size, the positive-gain region is sharply localized in $R_{\text{eff}}/\lambda<5$, where the ND beam wins $81.7\%$ of trials with a mean $\Delta P_{\text{rx}}=+6.13$~dB. The win rate falls below break-even already in the $5\le R_{\text{eff}}/\lambda<10$ bucket ($34.7\%$, $-2.36$~dB) and continues to drop monotonically for larger obstacles ($15.2\%$ in $[10,15)$ and $13.6\%$ in $[15,20)$), consistent with the headroom mechanism of Proposition~1: larger $R_{\text{eff}}$ inflates $z_{\min}$ and shrinks $H$ at the evaluated depths. Stratifying by the normalized depth $t$ instead, the per-bucket win rate stays in the $66\%$--$68\%$ band across $t\in[0,0.5]$, peaks at $68.4\%$ in $t\in[0.3,0.4)$, and then decays smoothly to $57.2\%$ at $t\in[0.7,0.8)$, $54.1\%$ at $t\in[0.8,0.9)$, $47.8\%$ at $t\in[0.9,1.0)$, and $30.4\%$ for $t\ge 1$ where the boresight reference has formally taken over the on-axis intensity (cf.~the definition of $z_{\text{crossover}}$ in Section~IV-A).

Fig.~\ref{fig:stats} reports the per-trial dispersion behind these aggregates over the full eight-family ensemble (the $64{,}680$ finite-extent trials together with the $9{,}240$ WallEdge half-plane trials, $73{,}920$ in total). The interquartile range is wide because the ensemble spans both the small-obstacle regime where ND self-healing dominates and the large-obstacle regime where the boresight reference benefits from edge diffraction (Poisson-Arago effects). The aggregate ND-vs-boresight win rate over the full ensemble is $59.3\%$ with a mean $\Delta P_{\text{rx}}=+1.85$~dB; the $(R_{\text{eff}},t)$ stratification above makes clear that this aggregate is the average of a strongly bimodal performance landscape rather than a uniform improvement.

\begin{figure}[!t]
    \centering
    \includegraphics[width=\columnwidth]{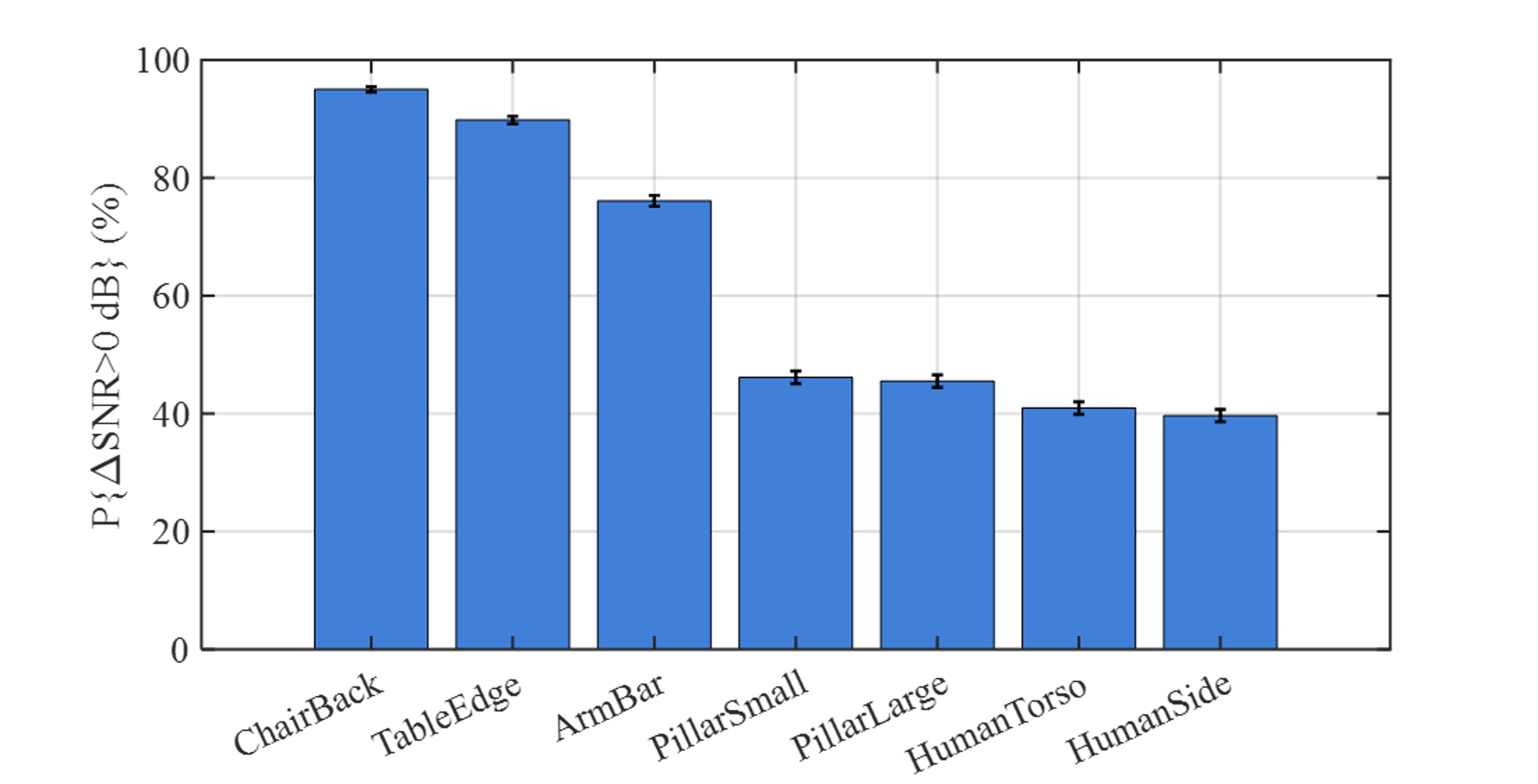}
    \caption{Per-scenario advantage probability of the isotropic Bessel-like beam against the boresight reference across the eight blocker families, aggregated over all trials with $t<0.95$. Error bars are $95\%$ Wilson confidence intervals.}
    \label{fig:scenarios}
\end{figure}

\begin{figure*}[!t]
    \centering
    \includegraphics[width=\textwidth]{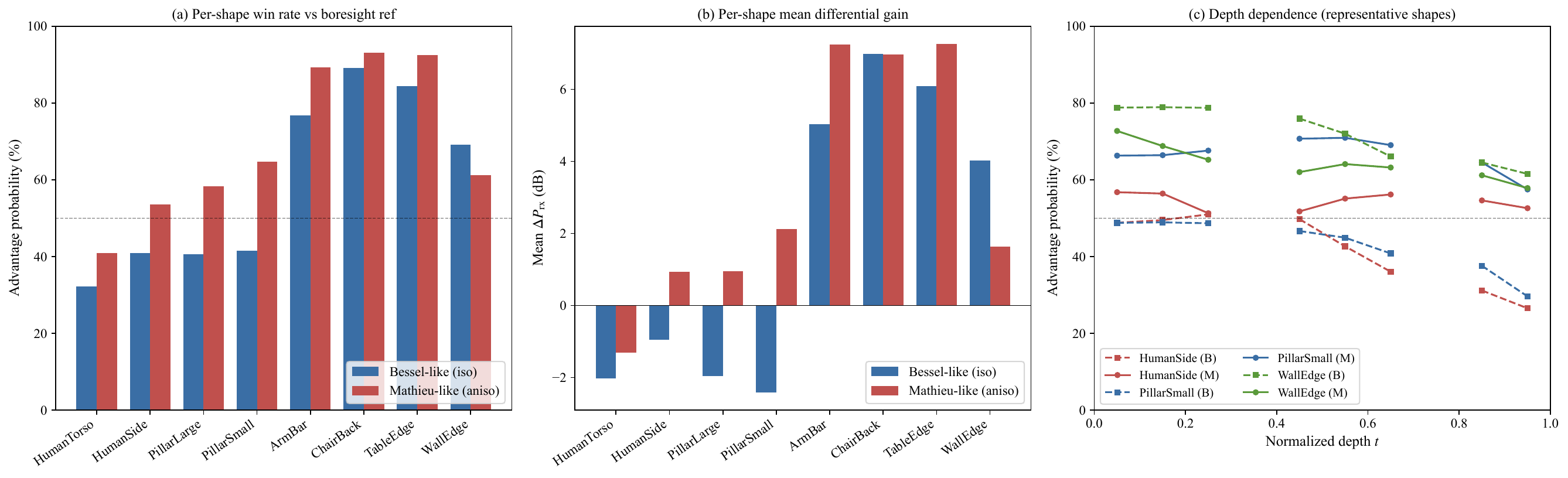}
    \caption{Anisotropic Mathieu-like versus isotropic Bessel-like performance. (a) Per-shape advantage probability (win rate) against the boresight reference; the eight shapes are grouped as the four vertical-sector weak spots (left) and the four remaining blockers (right). (b) Per-shape mean differential gain $\Delta P_{\text{rx}}$. (c) Advantage probability versus normalized depth $t$ for three representative shapes (HumanSide, PillarSmall, WallEdge); solid lines denote the Mathieu-like configuration, dashed lines the Bessel-like baseline.}
    \label{fig:mathieu}
\end{figure*}

\subsection{Mathieu-Like Geometry Adaptation}
While the aggregate statistics of Section~VI-B indicate a broad advantage region, decomposing the results by blocker family reveals a clear dependence on shape (Fig.~\ref{fig:scenarios}). The isotropic Bessel-like beam maintains a high win rate on horizontally extended or compact blockers (ChairBack $89.1\%$, TableEdge $84.5\%$, ArmBar $76.9\%$, WallEdge $69.1\%$) but performs substantially worse on vertically extended ones: HumanTorso $32.2\%$, HumanSide $41.0\%$, PillarLarge $40.7\%$, and PillarSmall $41.5\%$. These four low-performing shapes share a common feature: they intercept a predominantly vertical angular sector $\Phi_{\text{blk}}$ in the transverse $\bm{k}$-plane. Observation~2 indicates that an anisotropic spectrum whose high-density lobes are rotated away from $\Phi_{\text{blk}}$ should mitigate this deficiency.

We retain the same $64\times 64$ aperture, total radiated power, and propagation operator as in Section~III, and replace the isotropic spectral target with an anisotropic Mathieu-like target ($k_a>k_b$ with $\epsilon_2>0$ in \eqref{eq:target_spectrum}), concentrating $|F_{\text{target}}|$ along the horizontal axis so that its high-density lobes lie outside the vertical angular sector $\Phi_{\text{blk}}$. Fig.~\ref{fig:mathieu} reports the per-shape advantage probability under the same Monte Carlo blockage ensemble. All four vertical-sector shapes improve: HumanSide $41.0\%\to 53.6\%$, HumanTorso $32.2\%\to 41.0\%$, PillarLarge $40.7\%\to 58.3\%$, and PillarSmall $41.5\%\to 64.7\%$, with the largest gain ($+23.2$ percentage points) on the shape whose $\Phi_{\text{blk}}$ aligns most closely with the original Bessel weak axis. The high-baseline shapes also improve (ArmBar $76.9\%\to 89.2\%$, TableEdge $84.5\%\to 92.6\%$, ChairBack $89.1\%\to 93.0\%$), indicating that the orientation degree of freedom does not degrade the cases already handled well by the isotropic beam. The complete per-shape win rates and mean differential gains are given in Table~\ref{tab:mathieu_vs_bessel}.

\begin{table}[!t]
\centering
\caption{Per-Shape Advantage of Isotropic Bessel-Like vs.\ Anisotropic Mathieu-Like Beams\tnote{a}}
\label{tab:mathieu_vs_bessel}
\renewcommand{\arraystretch}{1.15}
\begin{threeparttable}
\begin{tabular}{@{}l c c c c c@{}}
\hline
 & \multicolumn{2}{c}{\textbf{Win rate (\%)}} & & \multicolumn{2}{c}{\textbf{Mean $\Delta P_{\text{rx}}$ (dB)}} \\
\cline{2-3}\cline{5-6}
\textbf{Blocker family} & Bessel & Mathieu & $\Delta$ pp & Bessel & Mathieu \\
\hline
HumanTorso  & 32.2 & 41.0 & $+8.8$  & $-2.03$ & $-1.31$ \\
HumanSide   & 41.0 & 53.6 & $+12.7$ & $-0.95$ & $+0.94$ \\
PillarLarge & 40.7 & 58.3 & $+17.6$ & $-1.97$ & $+0.94$ \\
PillarSmall & 41.5 & 64.7 & $+23.2$ & $-2.42$ & $+2.12$ \\
ArmBar      & 76.9 & 89.2 & $+12.4$ & $+5.03$ & $+7.25$ \\
ChairBack   & 89.1 & 93.0 & $+4.0$  & $+6.98$ & $+6.98$ \\
TableEdge   & 84.5 & 92.6 & $+8.1$  & $+6.09$ & $+7.27$ \\
WallEdge    & 69.1 & 61.3 & $\mathbf{-7.8}$ & $+4.03$ & $+1.63$ \\
\hline
\end{tabular}
\begin{tablenotes}
\item[a] Each row aggregates $9{,}240$ Monte Carlo trials ($21$ size scales $\times$ $11$ evaluation depths $\times$ $40$ random placements) under identical aperture, radiated power, propagation operator, and blockage realization. The $\Delta$ column is in percentage points. The single regression (WallEdge) corresponds to the half-plane geometry discussed in Section~VI-C.
\end{tablenotes}
\end{threeparttable}
\end{table}

The single regression is WallEdge ($69.1\%\to 61.3\%$). WallEdge is the only family in the library modeled as an infinite half-plane rather than a finite-extent blocker, so the intercepted angular sector $\Phi_{\text{blk}}$ is itself a half-plane: any anisotropic $|F_{\text{target}}|$ that concentrates spectral mass on one side of the cut still loses approximately half of its amplitude, and the pairwise form of the decision rule selects the isotropic Bessel-like configuration on this geometry. The $-7.8$ percentage-point reversal is therefore a boundary case, indicating that the orientation mechanism of Observation~2 is geometry-\emph{adaptive} and that its effect vanishes when the intercepted angular sector is itself unbounded. Overcoming this limit requires beam families whose transverse trajectory bends with propagation rather than being confined to a fixed $k_z$ cone; trajectory-bending beams of the Airy class are one such extension and have been shown to be realizable under the same phase-only phased-array constraint~\cite{qin2026airy,droulias2025}.

Two caveats accompany the absolute numbers. First, switching from the isotropic to the anisotropic spectral target slightly modifies the principal-axis cone angle $\theta_c$, and therefore the admissibility factor $H$ defined in \eqref{eq:Hfactor} is not held strictly constant between the two configurations; what is held constant is the aperture, the radiated power, and the evaluation window. Second, the figures above are aggregated over $t\in[0,1]$ and the rotation/offset distribution described in Section~V; per-depth behavior is reported in Fig.~\ref{fig:mathieu}. Even with these caveats, the dominant trend---simultaneous correction of the four vertical-sector weak spots---is consistent with the orientation mechanism of Observation~2.

\begin{figure}[!t]
    \centering
    \includegraphics[width=\columnwidth]{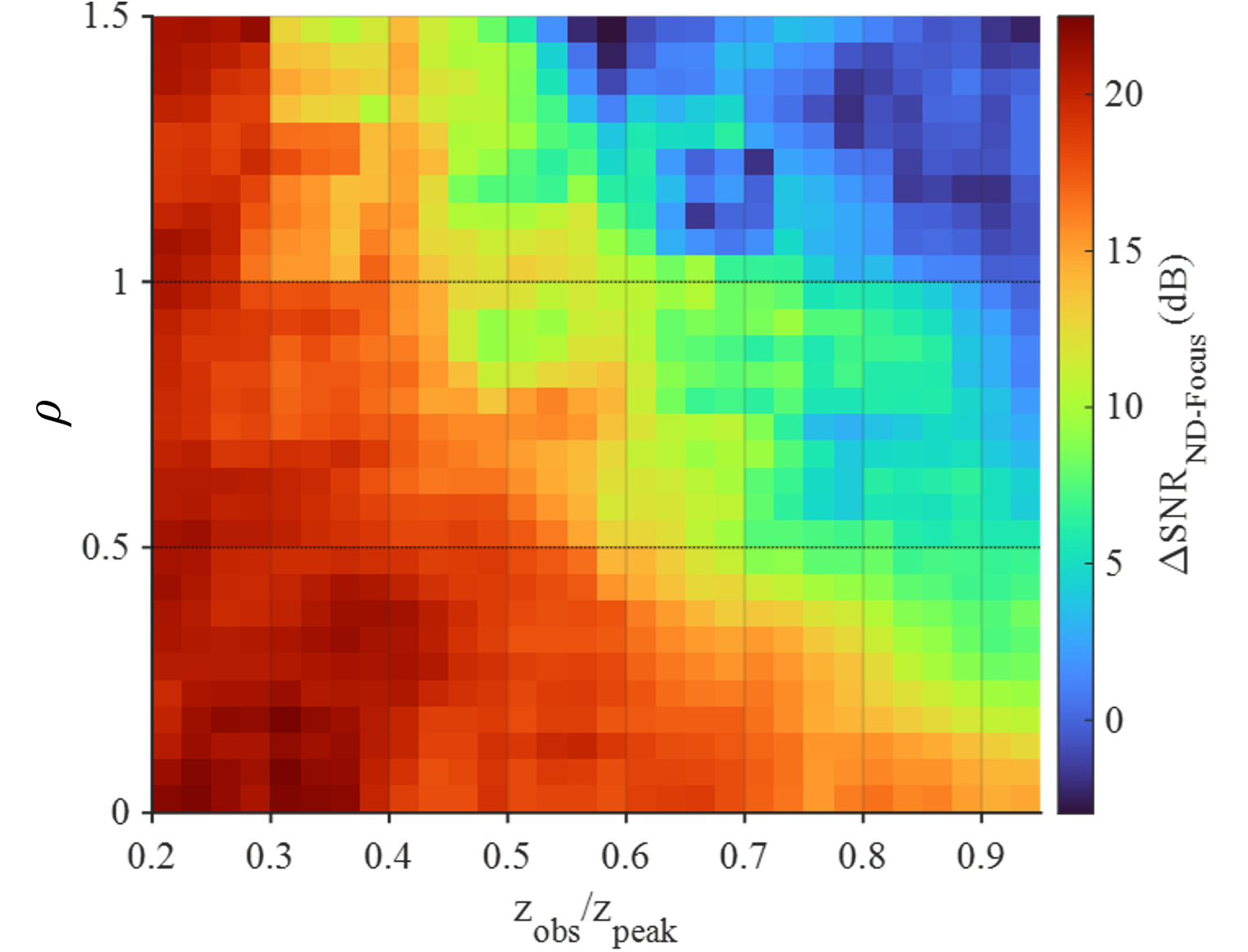}
    \caption{Differential gain $\Delta P_{\text{rx,ND-Focus}}$ of the ND beam over the near-field focusing (NF-F) baseline at $z_{\text{eval}}=z_{\text{peak}}$ ($t=0$). Axes are the normalized obstacle position $z_{\text{obs}}/z_{\text{peak}}$ and the difficulty index $\rho$.}
    \label{fig:nff}
\end{figure}

\subsection{Comparison with Conventional Near-Field Focusing}
The reference beam used throughout Sections~V and~VI-A--C is a uniform-phase, equal-aperture array baseline, chosen to isolate the contribution of the ND phase structure against a phase-flat control. A natural question is whether the ND advantage survives when the baseline is itself allowed to use its phase budget, specifically when it is compared against a near-field focusing (NF-F) array whose aperture phase is optimized to maximize the unblocked on-axis gain at the user coordinate. Fig.~\ref{fig:nff} reports that comparison, conducted under the same $64\times 64$ aperture, total radiated power, and propagation operator as the rest of the paper. Because the NF-F baseline concentrates all of its spatial degrees of freedom on the focal plane, the meaningful comparison point is the ND peak-intensity plane $z_{\text{eval}}=z_{\text{peak}}$ ($t=0$); the comparison at this single depth is therefore conducted across the full eight-family blockage ensemble, $8\times 21\times 40 = 6{,}720$ trials per beam.

Across the full ensemble, the ND beam achieves a positive advantage in $83.4\%$ of trials with a median improvement of $+11.4$~dB and a $[p_{10},p_{90}]$ interval of $[-2.7,+21.1]$~dB. Stratifying by the blockage difficulty index $\rho$ defined in \eqref{eq:rho_difficulty} shows a monotone decay along this coordinate: the median ND-vs-NF-F gain is $+15.9$~dB for low-difficulty obstacles ($\rho<0.2$, win rate $95.8\%$), $+10.9$~dB for moderate difficulty ($0.2\le\rho<0.5$, win rate $87.6\%$), $+8.0$~dB for hard obstacles ($0.5\le\rho<1.0$, win rate $77.3\%$), and only $+2.0$~dB in the high-difficulty regime ($\rho\ge 1.0$, win rate $59.6\%$). The trend is broadly consistent with the headroom mechanism of Proposition~1---larger $\rho$ corresponds to larger $z_{\min}/z_{\text{obs}}$ and, for fixed $z_{\text{eval}}$, smaller $H$---with the win rate falling smoothly toward $50\%$ rather than collapsing discontinuously.

Two observations follow. First, the comparison is consistent with the physical trade-off picture: the NF-F baseline trades blocked-link resilience for unblocked focal gain, and within the geometrically admissible regime this trade tends to favor the ND beam. Second, the comparison is not a uniform $15$~dB advantage---the $+15$~dB figure is reached only by the low-difficulty median---so the deployment value of ND beams over NF-F is regime-specific and quantified rather than asserted. The ND advantage reported here persists even against an unblocked-optimal focusing baseline evaluated at its own focal plane $z_{\text{eval}}=z_{\text{peak}}$.

\subsection{Robustness to the Cone-Angle Choice}
Proposition~1 predicts that the admissibility factor $H$ depends on the cone half-angle through $\cot\theta_c$, so the advantage window is expected to shift as $\theta_c$ changes rather than appear only at the single value used in Section~VI-B. We verify this explicitly by resynthesizing the Bessel-like beam at $\theta_c\in\{5^\circ,\,7^\circ,\,9^\circ\}$ under the same aperture, power, and propagation operator. The resulting unblocked crossover distances are $186$, $162$, and $137\,\lambda$ respectively: the window contracts with increasing $\theta_c$ in the direction predicted by $H$, and is non-empty at every angle tested. To verify that the advantage persists inside the rescaled window rather than only at its boundaries, we run a companion Monte Carlo ensemble of $200$ random circular-disk blockers per cone angle ($R_{\text{block}}\in[1,5]\,\lambda$, depth $z_{\text{obs}}\in[0.2,0.8]\,z_{\text{peak}}$, lateral offset $\in[-3,3]\,\lambda$), evaluated at $z_{\text{eval}}=z_{\text{peak}}$. The ND-vs-boresight win rate is $93.5\%$, $100\%$, and $100\%$ at $\theta_c=5^\circ$, $7^\circ$, and $9^\circ$, with mean differential gains of $+8.9$, $+10.2$, and $+14.5$~dB. The advantage is therefore positive and consistent in sign across the cone-angle range considered; the $(\rho,t)$-plane regime map in Fig.~\ref{fig:macro_map} rescales with $\theta_c$, and the qualitative beam-selection behavior of Sections~VI-A--D is preserved across this range.

\section{Validity, Scope, and Limitations}

\subsection{Fresnel Propagation and ASM Validity}
All propagation in this work is carried out with the exact angular-spectrum operator $\mathcal{P}_z$ defined in Section~III, which is valid for all propagating plane-wave components without a paraxial approximation. The relevance of the Fresnel / near-field regime enters only through the spatial scale at which the analysis is conducted. For an effective aperture radius $a$ and wavelength $\lambda$, the Fresnel number
\begin{equation}
F(z)=\frac{a^2}{\lambda z}
\end{equation}
measures near-field curvature. With the $32\lambda$ aperture used throughout ($a\approx 16\lambda$), the Fraunhofer distance is $z_{\text{FF}}\approx 2D^2/\lambda\approx 2048\lambda$, and every evaluation distance considered in Sections~V–VI satisfies $z<z_{\text{FF}}$ with $F(z)\gtrsim 0.1$, placing the advantage window squarely inside the radiating near field.

As an analytical sanity check, expanding the exact longitudinal phase $\sqrt{k_0^2-k_t^2}=k_0-k_t^2/(2k_0)-k_t^4/(8k_0^3)+\mathcal{O}(k_t^6)$ and bounding the neglected quartic term gives a simple wide-angle phase-error estimate for a paraxial surrogate,
\begin{equation}
\Delta\phi_{\max}(z)\lesssim \frac{\pi z}{4\lambda}\sin^4\theta_{\max}.
\end{equation}
For the maximum cone half-angle $\theta_{\max}\approx 7^\circ$ used in our ND designs, $\Delta\phi_{\max}\lesssim 1.7\times10^{-4}\,(z/\lambda)$ rad, i.e., below $2\times10^{-2}$ rad at $z=100\lambda$. This bound is reported for context only; the results themselves use the full non-paraxial ASM propagator and the thin-screen obstacle model introduced in Section~III.

\subsection{Per-Polarization Interpretation Under the Scalar Opaque-Screen Model}
This subsection clarifies the scope of the adopted scalar, polarization-agnostic formulation. Within the model, the angular-spectrum propagator $\mathcal{P}_z$ of \eqref{eq:transfer_func}, the synthesized aperture phase profile $\phi_b$ for $b\in\{\text{ND},\text{REF}\}$, and the binary obstacle mask $M(x,y)$ are all polarization-independent by construction. For a linearly polarized transmit aperture along any fixed unit vector $\hat{\mathbf{u}}$, each Cartesian component of the radiated field satisfies the scalar Helmholtz equation in homogeneous free space, and the mask extinguishes both components pointwise and identically. The reported win rates, landmark distances, and advantage ratios therefore apply to any fixed linear transmit polarization without modification, and the comparison can be interpreted on a per-polarization basis within the present model.

This statement is confined to the adopted model and should not be extrapolated to arbitrary vector obstacle interactions. For a real, partially transmissive, or edge-structured blocker, the obstacle response is governed by angle-dependent Fresnel coefficients, depolarization at edges, and surface-impedance effects. Because the ND beam and the reference beam illuminate the obstacle with different local angular spectra, these vector effects can weight the two arms asymmetrically and, in principle, perturb both absolute gains and relative trends. The results of this paper should therefore be interpreted as the geometry-driven component of the blocked-link comparison under an idealized opaque-screen obstacle; a full-wave vector treatment of specific materials and edge geometries is complementary to this scalar design-rule analysis.

\subsection{Frequency Dispersion and Wideband Operation}
The annular-spectrum target in \eqref{eq:target_spectrum} is defined through $k_\rho = k_0 \sin\theta_c$ and is synthesized at the carrier $f_c$. For fractional bandwidth $B/f_c\ll 1$ the cone angle and the landmark distances drift by $\mathcal{O}(B/f_c)$, so the $(\rho,t)$-plane advantage maps of Section~VI remain valid up to a corresponding linear rescaling; in the evaluated configurations ($B/f_c\sim 1$--$5\%$), these drifts remain within the advantage window. For OFDM operation the synthesis can be repeated per subcarrier (or per subcarrier group) without otherwise altering the analysis; true-time-delay arrays realize this natively, and phase-shifter arrays approximate it up to the beam-squint bound just cited.

\subsection{Partial-Transmission Blockers}
\label{ssec:alpha}
The opaque thin-screen model of Section~III is the limiting case $\alpha=0$ of a one-parameter family of partial-transmission masks $M(x,y)\in\{\alpha,1\}$, in which the interior of the blocker attenuates the incident field by a factor $\alpha$ rather than removing it entirely. This family embeds the opaque-limit advantage numbers of Section~VI into a one-parameter robustness study, and the sign of their variation with $\alpha$ is physically informative. Table~\ref{tab:alpha_sensitivity} reports the differential gain $\Delta P_{\text{rx}}$ for an isotropic Bessel-like beam against the boresight reference over $\alpha\in\{0,\,0.1,\,0.3\}$ on five representative blocker shapes, with the obstacle at $z_{\text{obs}}=0.6\,z_{\text{peak}}$ and the user at $z_{\text{eval}}=z_{\text{peak}}$.

\begin{table}[!t]
\centering
\caption{Differential Gain $\Delta P_{\text{rx}}$ (dB) Under a Partial-Transmission Mask $M\in\{\alpha,1\}$, Bessel-Like vs.\ Boresight Reference at $z_{\text{eval}}=z_{\text{peak}}$\tnote{a}}
\label{tab:alpha_sensitivity}
\renewcommand{\arraystretch}{1.15}
\begin{threeparttable}
\begin{tabular}{@{}l c c c@{}}
\hline
\textbf{Shape} & $\alpha=0$ (opaque) & $\alpha=0.1$ & $\alpha=0.3$ \\
\hline
Compact disk ($R=3\lambda$)     & $+6.65$ & $+7.24$  & $+8.30$  \\
Vertical bar ($2\times 20\lambda$) & $+9.97$ & $+10.08$ & $+10.23$ \\
Half-plane ($x>-2\lambda$)      & $+9.69$ & $+9.81$  & $+9.99$  \\
HumanTorso-shaped ($4\times 16\lambda$) & $+5.44$ & $+6.33$ & $+7.80$ \\
PillarLarge ($R=5\lambda$)      & $+6.60$ & $+8.67$  & $+12.90$ \\
\hline
\end{tabular}
\begin{tablenotes}
\item[a] $z_{\text{obs}}=0.6\,z_{\text{peak}}$, $z_{\text{eval}}=z_{\text{peak}}$. Each entry is a single deterministic propagation with the indicated blocker centered on the optical axis. The consistent sign and trend across all five shapes and both nonzero $\alpha$ values indicate that the qualitative conclusion is robust; an exhaustive Monte Carlo validation across the full $73{,}920$-trial ensemble under partial transmission is a natural extension that does not affect the design-rule framing of this paper.
\end{tablenotes}
\end{threeparttable}
\end{table}

The trend in Table~\ref{tab:alpha_sensitivity} is initially counter-intuitive. As expected, the sign of $\Delta P_{\text{rx}}$ is preserved across every row; however, the magnitude is \emph{non-decreasing} in $\alpha$: partial transmission widens the ND-versus-reference gap on every shape tested, rather than closing it. The mechanism is physically analogous to the geometric-headroom picture of Proposition~1, but applied here to transmissive attenuation rather than to conical reconstruction distance. The boresight reference concentrates its unblocked energy along a narrow on-axis region that the obstacle shadows directly, so a nonzero $\alpha$ transmits energy into precisely the spatial modes on which the reference most depends. The ND beam, in contrast, draws its on-axis intensity from conical components that reach the receiver from directions only partly shadowed by the obstacle, and a nonzero $\alpha$ contributes to those conical components without canceling the on-axis term that dominates the reference. For the representative cases evaluated here, the opaque-screen model of Sections~V and~VI appears to be a \emph{conservative} reading of the advantage: partial transmission increases the measured differential gain rather than reducing it.

\section{Conclusion}
This paper has developed a geometry-aware framework for evaluating non-diffracting beams as a transmitter-side blockage mitigation in near-field mmWave links. The framework centers on a necessary admissibility condition ($H>1$) that links the blocker geometry to the usable non-diffracting range, an orientation-based degree of freedom for anisotropic Mathieu-like spectra, and a link-level beam-selection rule that maps to an achievable-rate gap across the operating SNR range.

Monte Carlo evaluation across $73{,}920$ blocked-link trials with eight blocker families shows that the isotropic Bessel-like beam wins $59.3\%$ of the ensemble ($\Delta P_{\text{rx}}=+1.85$~dB mean), with the advantage concentrated in the low-difficulty geometric regime and decaying as $H$ approaches unity. The anisotropic Mathieu-like beam improves the four vertically extended blocker families by $+8.8$ to $+23.2$ percentage points without degrading the remaining shapes, while the half-plane WallEdge case---where no azimuthal redistribution can compensate an unbounded angular shadow---marks the structural boundary of the annular-spectrum approach. An auxiliary $6{,}720$-trial comparison against an unblocked-optimal near-field focusing baseline at $z_{\text{peak}}$ yields a positive ND advantage in $83.4\%$ of trials ($+11.4$~dB median), and robustness studies over cone-angle choice and partial-transmission blockers confirm that the advantage is preserved across the evaluated design space.

Trajectory-bending beam families, which act on a degree of freedom orthogonal to the orientation mechanism analyzed here, are a natural complement to the present framework and are left to future work~\cite{qin2026airy}.

\end{document}